\documentstyle[11pt,newpasp,twoside,epsf]{article}
\def\xmm{{\em XMM}}

\def\ein{{\em Einstein}}
\def\ros{{\em ROSAT}}
\def\asc{{\em ASCA}}
\def\cha{{\em Chandra}}

\def\hea{{\em HEAO-1}}

\def\flux{erg cm$^{-2}$ s$^{-1}$}
\def\sdeg{deg$^{2}$}
\def\~{$\sim$}

\def\edcomment#1{\iffalse\marginpar{\raggedright\sl#1\/}\else\relax\fi}
\reversemarginpar

\begin{document}

\title{X-ray Cluster Large Scale Structure and Cosmology}
 \author{Marguerite PIERRE}
\affil{Service d'Astrophysique,  CEA Saclay, F-91191 Gif Sur
Yvette}

\begin{abstract}
We outline  main arguments in favor of cosmological X-ray surveys
of galaxy clusters. We summarize recent advances  in our
understanding of cluster physics. After a short review of past
surveys, we present the scientific motivations of the XMM Large
Scale Structure survey. We further illustrate how such a survey
can help constrain  the nature of the dark energy as well as
cluster scaling law evolution, i.e. non gravitational physics.
\end{abstract}

\section{Introduction}

Clusters of galaxies occupy a strategic position in the
multi-parameter cosmological space:

- They constitute the most massive entities having reach some
equilibrium state in the universe. Masses range from a few
10$^{13}$ $M_{\odot}$ for groups up to a few 10$^{15}$ $M_{\odot}$
for rich Coma-type clusters;

- They are located at the nodes  of the cosmic network.

- Maps of the distribution of clusters trace the matter
distribution over large volumes;

- Clusters grow from accretion at a rate that depends on the
embedding cosmology;

- As ``relaxed'' objects, they can be considered to have decoupled
from the general expansion;

- In the cluster total mass budget, galaxies account for only 5\%,
intra-cluster gas for 15\%, the rest of the mass being in the form
of
 dark matter;

- ``Equilibrium'' is twofold: virial equilibrium for the galaxies
and hydrostatical equilibrium for the gas within the cluster
potential;

\medskip

\noindent This short list provides compelling arguments for using
clusters as cosmological probes rather than galaxies being
complex, highly non-linear objects. We should, however, underline
that the physics of clusters, as revealed by the two last
generations of X-ray observatories (\ros\ and \asc\, then \xmm\
and \cha), is not as simple as suggested by the above picture:
phenomena other than gravity also impact on cluster properties.
They are mostly  triggered by interactions  involving gas and
(active) galaxies within the dark matter potential. Among the
currently investigated questions, we may cite:

- Cluster relaxation processes and time scales after merger
events, i.e. resorption of clumps. Observables involved are:
galaxy velocity and space distribution; gas density and
temperature maps; radio halos.

- Modelling gas cooling  and understanding the cooling flow
problematic. This involves extensive numerical simulations as well
as high resolution X-ray spectroscopy;

- The presence of magnetic fields and of a  potentially non
negligible contribution from high velocity electrons (cosmic rays)
in the total cluster energetic budget;

- The impact of stellar activity (supernovae) in cluster galaxies
to the heating and enrichment of the cluster gas;

- The role of gas stripping and accretion in the evolution of
cluster galaxies;

- The role of cluster AGN in the energetic budget and magnetic
field properties. This is especially relevant for cooling flow
studies;\\
A review of these   aspects  can be found in Mushotzky (2004, this
volume).

\section{Why X-ray cluster surveys?}

Because of the many links with fundamental physics, X-ray cluster
data offer significant advantages over optically-selected cluster
samples.

 In absence of extra heating (other than shocks) and
cooling mechanisms, the gas  trapped in the cluster gravitational
potential is heated up to the virial temperature:
$$ kT \sim 6.7 ~(\frac{M_{200}}{10^{15}M_{\odot} h^{-1}})^{2/3} ~~ keV$$
where $M_{200}$ is the mass enclosed within the virial
radius\footnote{The virial radius is usually defined as
cluster-centric radial distance where the dark matter density is
200 times higher than that of the mean critical density considered
at the cluster's redshift}. From this, it occurs that cluster
temperatures approximatively range from 1 to 10 keV which
correspond to wavelengths of the order of 10-1 \AA, i.e. to the
X-ray domain. The density of the intra-cluster medium is of the
order of one atom per liter, its metallicity about 0.3 solar, most
of the gas being totaly ionized, except for some heavy elements
like iron. The X-ray emission from such an optically thin plasma
can be described by a bremsstrahlung continuum emission, plus
possible fluorescence lines from heavy elements. The emissivity is
thus simply proportional to the square of the electron
density. \\
From this, one expects  simple scaling relations   connecting
cluster X-ray luminosities to temperatures and, further, to
cluster masses (the only parameter that enter any cosmological
consideration). Observations do indeed show strong correlations
between $L-T-M$, but with a relatively large scatter. Although the
dispersion can be ascribed, for a large part, to the individual
cluster formation and relaxation histories,  the relations provide
useful tools to link observations to  theory in statistical
analyses. Moreover, the fact that the slopes of these relations
were found to be somewhat steeper than that predicted from simple
scaling laws implied by the virial equilibrium, pointed toward
additional heating (or cooling) mechanisms, other than purely
gravitational. Main mechanisms usually invoked are mentioned in
the previous section and their relative efficiency is investigated
by means of hydrodynamical simulations. \\

From the observational point of view, the presence of extended
X-ray emission at high galactic latitude almost unambiguously
points toward a deep cluster potential well. Moreover, projection
effects are much less of a concern than in the optical. At
moderate sensitivity ($\sim 10^{-14}$ erg/s/cm$^{2}$ in the
[0.5-2] keV band, which is obtained in a few ks with \xmm), the
X-ray sky is ``clean''. With a source density of about 200
$deg^{-2}$, clusters represent some 15\% of the population,  the
rest being mostly point-like active galactic nuclei. For
comparison, an extragalactic field (so-called ``empty field'')
observed in the I optical band in one hour with a 4m-class
telescope reveals a faint galaxy density of the order of $10^{5}~
deg^{-2}$. Sophisticated multiresolution wavelet-based algorithms,
now offer reliable means to flag the presence of faint extended
sources down to the (Poissonian) limit of the X-ray photon signal.
This renders the detection of X-ray clusters substantially more
direct and more quantitative than in the optical. Selection
effects can be
modelled by means of extensive simulations (Valtchanov et al, 2001). \\
Finally, because clusters are among the intrinsically brightest
X-ray sources, the X-ray domain is a priori ideally suited to
investigate the distant universe that is, well beyond $z=1$.

\section{Cluster surveys prior to XMM}

\hea\ (1977) was the first mission to provide an X-ray all-sky
survey enabling cluster statistical studies.   Some 100 nearby
clusters were inventoried. First determination of the cluster
X-ray luminosity function and a qualitative study of the sky
distribution of X-ray clusters, as well as correlations between
$L_{X}$ and optical richness were attempted
(Piccinotti et al, 1982 and  Johnson et al, 1983).\\
 With the advent of X-ray imaging
with focussing optics in the 80s, particularly with \ein, a new
era was opened in X-ray cluster surveys. The \ein\ Medium
Sensitivity Survey provided a sample of 93 clusters out to a
redshift of 0.58 and a flux limit of S$_{[0.3-3.5]~ keV}$ = 1.33
$10^{-13}$. These ``serendipitous'' sources were found in the
field of unconnected pointed observations covering a total of  780
$deg^{2}$ (Henry et al, 1992) and provided first hints about X-ray cluster evolution.\\
In 1990, the \ros\ All-Sky Survey  (RASS) was the first X-ray
imaging mission to undertake the coverage of the entire sky (Voges
et al, 1999). With average sensitivities  of the order of 2-20
10$^{-14}$ \flux\ in the [0.1-2.4] keV band, (FWHM \~ 100\arcsec)
and  the low instrumental background of the PSPC detector, the
RASS laid the basis for numerous unprecedented statistical
studies. After years of intensive follow-up campaigns, more than
1000 RASS clusters out to a redshift of $\sim 0.5$ were
inventoried. (1) Samples covering a contiguous area, thus suitable
for large scale structure studies. REFLEX for the southern
hemisphere with 450 objects ($z \leq  0.3$) it is the largest
homogeneous compilation to date, down to a flux limit of 3
10$^{-12}$ \flux\ in [0.2-2.4] keV (B\"ohringer et al, 2001); NORA
for the Northern hemisphere (B\"ohringer et al, 2002), and the
North Ecliptic Pole survey involving the 80 \sdeg\ deepest region
of the RASS provided 64 clusters out to $z \sim 0.81$ (Henry et
al, 2001). (2) Samples gathering well defined classes of clusters
such as the Massive Cluster Survey (Ebeling et 2001)  with the
goal of detecting the most massive clusters down to the RASS
sensitivity
limit.\\

After the completion of the RASS, \ros\ was run like an ordinary
observatory, performing thousands of deep targeted guest
observations. Many of these pointings  were suited to the search
for serendipitous clusters, following onto the EMSS heritage.
Numerous studies led to the detection of  several hundreds of
clusters out to a redshift to unity or  above. The large majority
of the serendipitous data indicate a mild cluster evolution of
cluster properties, compatible with a $\Lambda$CDM-type cosmology.
A summary of the survey work  and associated cosmological
constraints can be found  in Rosati et al, (2002).
We describe below in more detail the impact of LSS studies. \\

\section{Constraining cosmology}

In the current quest for the cosmological parameters, clusters
studies, beside CMB and supernova studies, provide critical
independent constraints as they do not rely on the same physical
phenomena. It is also necessary, as a consistency check, to
compare constraints obtained separately from the high and low
redshift universes. Like for galaxies, topological investigations
may involve tools as simple as the 2-point correlation function,
percolation analysis or more refined ones like Minkowski
functionals or the genus approach (e.g. Kerscher et al, 1997)
necessitating, however, rather high-level statistics. The cluster
correlation function which is proportional to the Fourier
transform of $P(k)$, the power spectrum of density fluctuations,
is quite sensitive to the spectrum shape ($\Gamma$). The amplitude
of $P(k)$ is strongly constrained by the cluster number density
and usually expressed in terms of $\sigma_{8}$, the r.m.s density
fluctuations within a top hat sphere of 8 h$^{-1}$  Mpc radius
(the local determination of cluster abundance, however, only
enables the determination some combination of $\sigma_{8}
\Omega_{m}^{\alpha}$). It is thus especially relevant to combine
both quantities in order to tighten the possible parameter space
(e.g. Moscardini et al, 2001). A didactic illustration of such a
procedure  can be found in Schuecker et al, (2003) for the REFLEX
sample.

As for any survey, constraining cosmological parameters requires
proper modelling of the selection effects, which is best achieved
with simulations. In the case of X-ray cluster surveys, special
attention must be paid to sensitivity variations across the survey
area and to the fact that some clusters may remain undetected
(e.g. low surface brightness objects, but not necessarily low-mass
entities). Further sources of uncertainty such as the intrinsic
scatter present in the $L_{X}-M$ relation must be integrated into
the final calculations. For high-z redshift samples,  evolutionary
considerations as to cluster scaling laws  have   also to be taken
into account.

\section{The XMM Large Scale Structure Survey}

Tracking back evolution in cluster properties largely improves
constraints on cosmological parameters, to which clusters are
especially sensitive like the matter density of the universe and
also, in some respect,  the nature of the dark energy (see below).
In particular, it allows breaking the $\sigma_{8}
\Omega_{m}^{\alpha}$ degeneracy (Bahcall et al, 1997).

Launched in 1999, the XMM observatory offers unrivalled collecting
area together with good imaging capabilities and a large field of
view. This translates into the following numbers: a PSF of
6\arcsec , which enables flagging clusters as extended sources out
to a redshift of 2, if any; a sensitivity reaching 5 $10^{-15}$
\flux\ in the [0.5-2] keV band for a 10 ks exposures and a field
of view of 30\arcmin . Although not initially conceived as a
survey instrument, its properties make of XMM an ideal cluster
finder. Taking advantage of this unique opportunity, we have
undertaken an extragalactic medium deep survey, the XMM Large
Scale Structure survey (XMM-LSS), with the goal of investigating
for the first time evolutionary trends in the space distribution
of clusters out to a redshift of unity. Requiring an accuracy for
the cosmological parameters comparable to that achieved by the
REFLEX low-$z$ sample, implies  obtaining some $2 \times 400$
clusters in the $0<z<0.5$ and $0.5<z<1$ redshift bins. This can be
obtained from a $8\times 8$ \sdeg\ area paved with 10 ks XMM
pointings (assuming a $\Lambda$CDM universe). Such a geometry also
ensures probing characteristic scales significantly larger than
100$h^{-1}$ at $z=1$. The X-ray survey with its associated
multi-wavelength programmes are described in detail by Pierre et
al, (2003). We  summarize below main expected cosmological
implications, especially in the context of the recent WMAP
results.

\section{Cosmological implications of the XMM-LSS}

Prospects that the final XMM-LSS catalogue will offer for
measuring cosmological parameters have been studied in detail by
Refregier et al, (2002) for a $\Lambda$CDM model. Cluster
abundance data provide strong constraints on the
$\Omega_{m}-\sigma_{8}$ combination. Adding information from the
correlation function restrains the allowed region of the
$\Omega_{m}-\Gamma$ plane to a narrow range. Given the survey
design, the $simultaneous$ expected precision on $\Omega_{m},
\sigma_{8}$ and $\Gamma$ is about 15\%, 10\%, 35\% respectively.

\begin{figure}
    \plotone{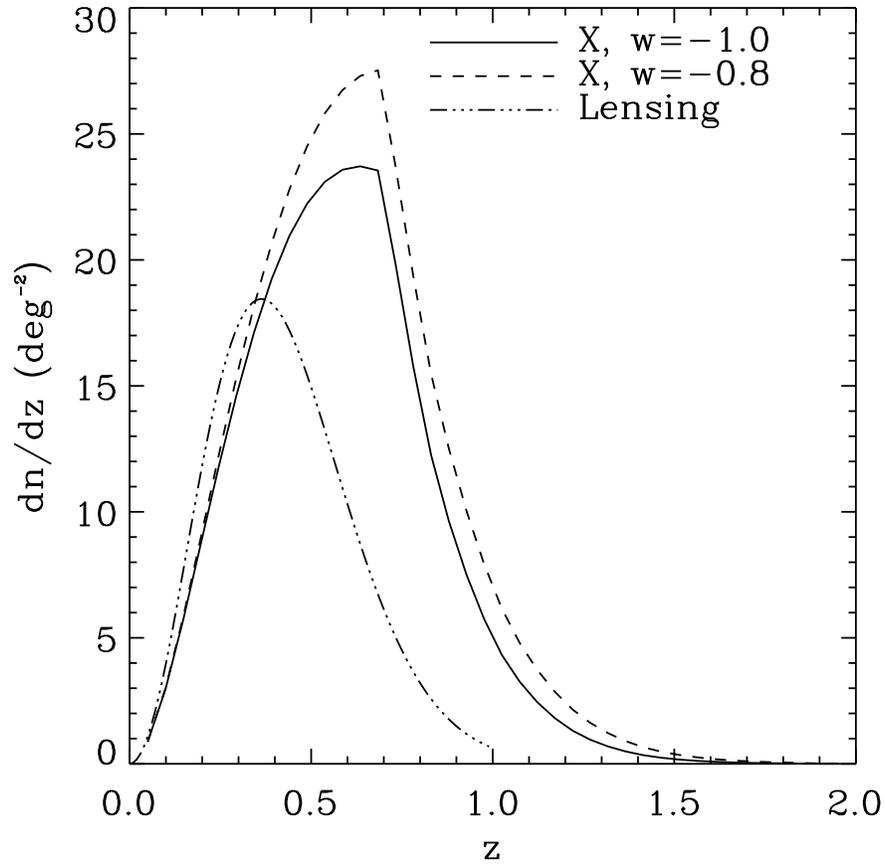}
    \caption{Predicted redshift distribution, $dn/dz$, for clusters selected
  in the X-ray with the XMM-LSS ($kT > 2 \ \rm keV$ and
  $S_{[0.5-2.0 \ {\rm keV}]} > 8 \times 10^{-15} \ \rm erg \, cm^{-2} \,
  s^{-1}$). The X-ray cluster counts are shown both for a $\Lambda$CDM
  model (with $w = -1$) and for a QCDM model ($w = -0.8$). We also show
  the predicted cluster counts expected from the weak lensing analysis
  (Bartelmann et al, 2002)}
\end{figure}

It is well known  that compared to $\Omega_{m}=1$ models, open
universes show a much less rapid evolution of the cluster number
density (see Refregier et al, 2002, for comparative $n(z)$ for the
XMM-LSS). We further show on Fig. 1 the predicted X-ray cluster
redshift distribution for two flavours of the dark energy.  X-ray
counts appear to be sensitive to the nature of the dark energy and
to be more sensitive for detecting clusters above $z>0.3$ than
ground-based weak lensing. In addition, thanks to the large area
surveyed, it will be possible to constrain the population of X-ray
bright (massive?) clusters above $z>1$. For a $\Lambda$CDM
cosmology, the probability of finding a Coma-type cluster (8 keV)
between $1.5<z<2$ over the entire survey area of 64 \sdeg\ is of
the order of 0.001, compared to 0.3 in the $0.5<z<1$ range;
getting a few such high-z clusters, would thus be most interesting!\\

Given the fact that we are now entering a high-precision cosmology
era with CMB and space supernovae experiments, we may address the
question of cluster evolution  from a different point of view.
Assuming that the cosmological parameters are known,  a cluster
survey can be used to constrain the evolutionary trends of the
scaling relations and, subsequently, characterize the impact of
the many processes presented in Sec. 1 \& 2 on cluster physics and
observables.

\begin{figure}
    \plottwo{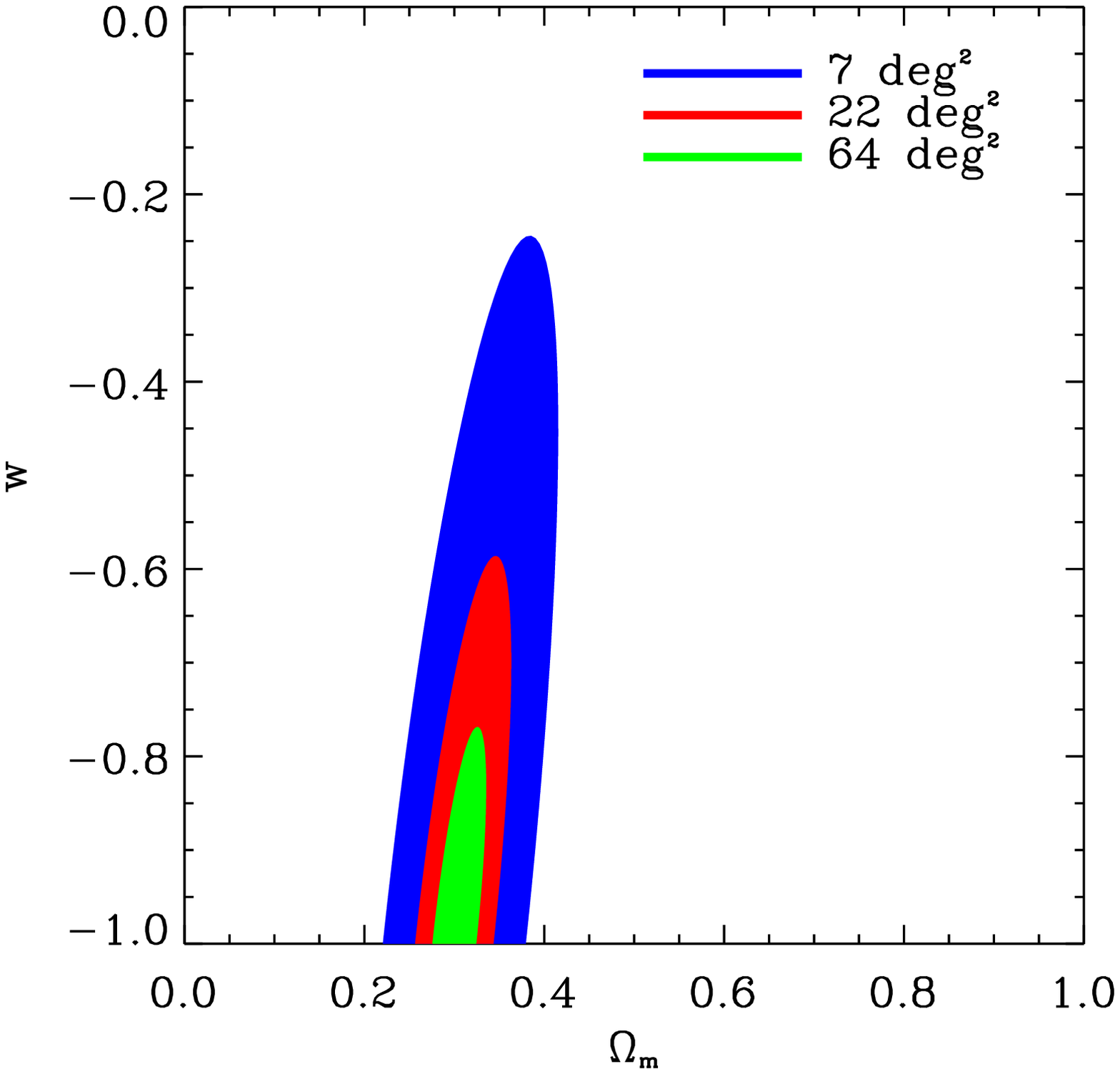}{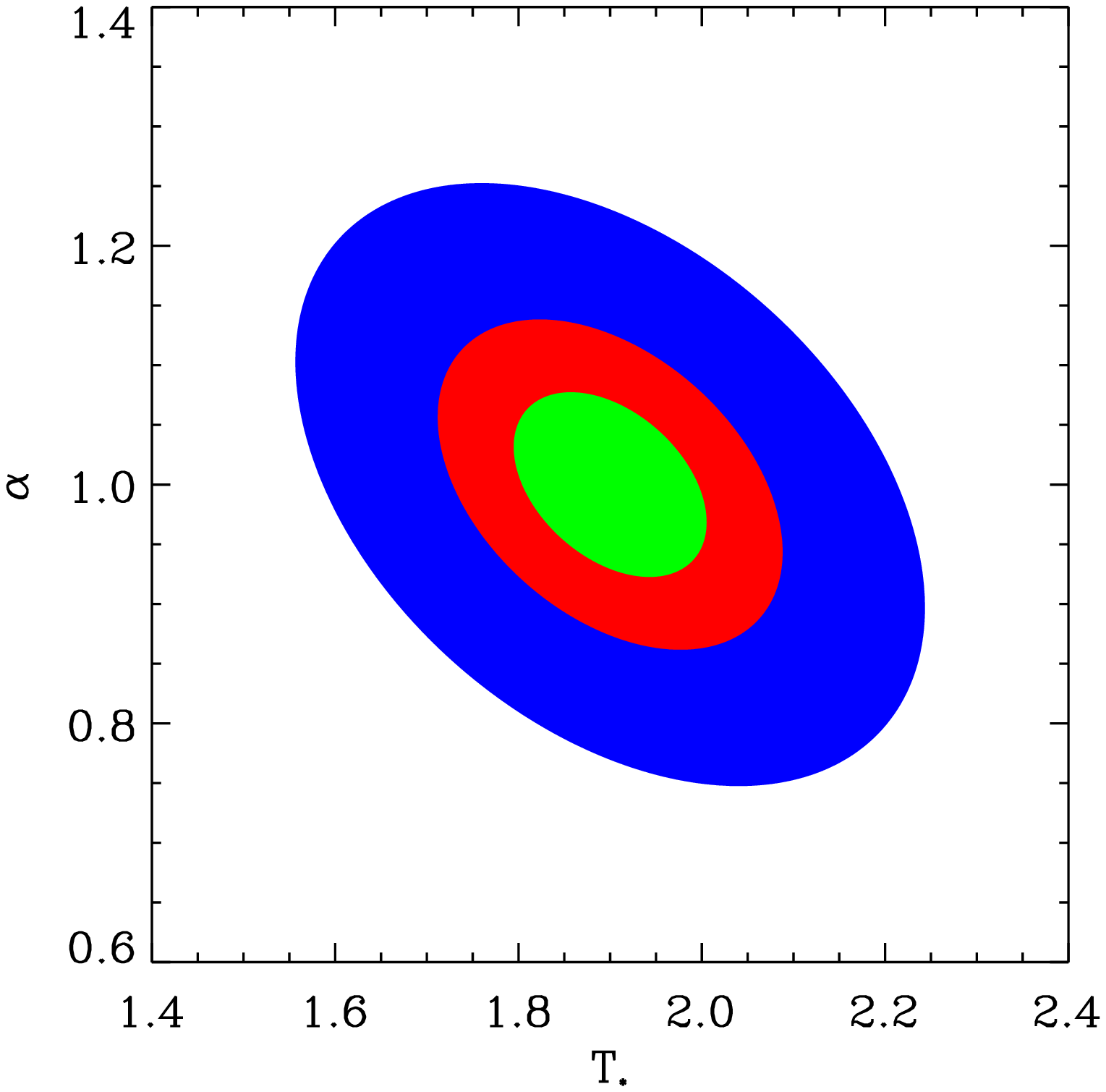}
\caption{ \underline{Left:} Constraints on cosmological parameters
 at various stage of XMM-LSS coverage. Priors for $\sigma_{8}$
  from WMAP and mean cluster scaling relations were assumed.
  \underline{Right:} An example of constraints on the cluster scaling
  relations from
  the survey. The cosmological parameters $\Omega_{\rm m}$ and
  $H_0$ were fixed to the WMAP values. The prior from WMAP for the
  $\sigma_{8}$ was assumed (a preliminary figure from Refregier et al, 2004).
  }
\end{figure}

Both aspects of such an approach are illustrated on Fig. 2 for
different stages of the survey coverage. The XMM-LSS
  survey could improve on WMAP's measurement of $\Omega_{m}$
  and  constrain the dark energy equation of
  state parameter  $w$, to which WMAP is not sensitive (Fig. 2, left).
  Alternatively, assuming that the cosmological parameters are
  well constrained (by a combination of WMAP, supernovae
  observations and nearby very large scale galaxy surveys),
  the evolution of the cluster scaling laws can be followed
  (Fig. 2 right). The parameters $T_\ast$
  and $\alpha$ describe the amplitude and evolution of the
  mass-temperature relation ($M \propto T^{3/2}$ for $T > 2 \ \rm
  keV$). The XMM-LSS has the potential of determining $T_\ast$ and resolving the
  controversy of its value, which has been given as low as $1.2$ or as
  high as $1.9$ (see Pierpaoli et al, 2003). The survey is also able to
  constrain the evolution parameter, $\alpha$, which has been
  fixed to 1 in earlier work. This is possible with the XMM-LSS survey because
  of the large cluster sample and wide range of redshifts.

  At a later stage,
  observations from associated   Sunayev-Zel'dovich and weak lensing surveys will yield
  further constraints on the scaling relation of clusters, and will
  impose   {\it simultaneous} constraints on cosmological parameters and
  cluster physics.

\end{document}